\documentclass{svmult}

\usepackage{mathptmx}       
\usepackage{helvet}         
\usepackage{courier}        
\usepackage{type1cm}        
\usepackage{makeidx}         
\usepackage{graphicx}        
\usepackage{multicol}        
\usepackage[bottom]{footmisc}

\usepackage{amsmath,amscd, amsfonts}

\newtheorem{thm}{Theorem}[section]

\newcommand{\Pin}{\mathop{\mathrm{Pin}}}
\newcommand{\Spin}{\mathop{\mathrm{Spin}}}

\usepackage{lineno}
\usepackage{tikz}
\usepackage{pgfplots}
\usepgflibrary{shapes.geometric} 
\usepgflibrary[shapes.geometric] 
\usetikzlibrary{shapes.geometric} 
\usetikzlibrary[shapes.geometric] 

\makeindex             

\begin{document}

\title*{A Clifford algebraic framework for  Coxeter group theoretic computations}
\author{Pierre-Philippe Dechant}
\institute{Pierre-Philippe Dechant \at Institute for Particle Physics Phenomenology \\Ogden Centre for Fundamental Physics \\ Department of Physics \\ University of Durham \\ South Road \\ Durham, DH1 3LE \\ United Kingdom \\ \email{pierre-philippe.dechant@durham.ac.uk}}
%
%
\maketitle

\abstract{Real physical systems with reflective and rotational symmetries such as viruses, fullerenes and quasicrystals have recently been  modeled  successfully in terms of three-dimensional (affine) Coxeter groups.
Motivated by this progress, we explore here the benefits of performing the relevant computations in a Geometric Algebra framework, which is particularly suited to describing reflections. 
Starting from the Coxeter generators of the reflections, we  describe how the relevant chiral (rotational), full (Coxeter) and binary polyhedral  groups can be easily generated and treated in a unified way in a versor formalism. 
In particular, this yields a  simple construction of the binary polyhedral groups as discrete spinor groups.
These in turn are known to generate Lie and Coxeter groups in dimension four, notably the exceptional groups $D_4$, $F_4$ and $H_4$.  
A Clifford algebra approach thus reveals an unexpected connection between Coxeter groups of ranks 3 and 4. 
We  discuss how to extend these considerations and computations to the Conformal Geometric Algebra setup, in particular for the non-crystallographic groups, and construct root systems and quasicrystalline point arrays. We finally show how a Clifford versor framework sheds light on the geometry of the Coxeter element and the Coxeter plane for the examples of the two-dimensional non-crystallographic Coxeter groups $I_2(n)$ and the three-dimensional groups $A_3$, $B_3$, as well as the icosahedral group $H_3$.
\\IPPP/12/49, DCPT/12/98\\
}

\section{Introduction}\label{sec_intro}

Physical systems have to obey the mathematical laws of geometry; in particular, if they possess symmetry -- such as invariance under reflections and rotations -- this symmetry is heavily constrained by purely geometric considerations. For instance, many physical systems in biology (viruses), chemistry (fullerenes) and physics (crystals and quasicrystals) have polyhedral symmetries. 
These polyhedral symmetry groups are generated by reflections; via the Cartan-Dieudonn\'e theorem an even number of reflections amounts to a rotation (see e.g. \cite{Garling2011Clifford} for an exposition in a Clifford algebra context), and  physical systems may be invariant only under this rotational (chiral) part, or the full reflection group.

 Coxeter group theory \cite{Coxeter1934discretegroups, Humphreys1990Coxeter} axiomatises reflections from an abstract mathematical point of view. 
Coxeter groups thus encompass the finite Euclidean reflection groups, which include the symmetry groups of the Platonic solids -- $A_3$ for the tetrahedron, $B_3$ for the dual pair octahedron and cube, and $H_3$ for the dual pair icosahedron and dodecahedron -- as well as the Weyl groups of the simple Lie algebras.
A subset of these groups are non-crystallographic, i.e. they describe symmetries that are not compatible with lattices in  dimensions equal to their rank. 
They include the two-dimensional family of symmetry groups $I_2(n)$ of the regular polygons, as well as $H_2$ (the symmetry group of the decagon), $H_3$ (the symmetry group of the icosahedron) and the largest non-crystallographic group $H_4$ (the symmetry group of the hypericosahedron or 600-cell in four dimensions), which are the only Coxeter groups generating rotational symmetries of order 5.   The full icosahedral group $H_3$ and its (chiral) rotational subgroup $I$ are of particular practical importance, as $H_3$ is the largest discrete symmetry group of physical space. Thus, many 3-dimensional systems with  `maximal symmetry', like  viruses in biology \cite{Stockley2010emerging, Caspar:1962, Twarock:2006b, Janner:2006b, Zandi:2004}, fullerenes in chemistry \cite{Kroto:1985,Kroto:1992, Twarock:2002b, Kustov:2008},  quasicrystals in physics \cite{Katz:1989, Senechal:1996, MoodyPatera:1993b, Levitov:1988, Shechtman:1984} as well as polytopes in mathematics \cite{Koca2007Polyhedra,Koca2010Catalan,Koca2010QuasiregularII}, can be modeled using Coxeter groups.

Clifford's Geometric Algebra \cite{Hestenes1966STA, LasenbyDoran2003GeometricAlgebra} is a complementary framework that 
focuses on the geometry of the physical space(-time)  that we live in and its given Euclidean/Lorentzian metric. This exposes more clearly the geometric nature of many problems in mathematics and physics. In particular, Clifford's Geometric Algebra has a uniquely simple formula for performing reflections. Previous research appears to have made exclusive use of one framework at the expense of the other. Here, we combine both paradigms, which results in geometric insights from Geometric Algebra that apparently have been overlooked in Coxeter theory thus far. This approach also has computational and conceptual advantages over standard techniques, in particular through a spinorial or conformal point of view. Hestenes \cite{Hestenes2002PointGroups} has given a thorough treatment of point and space groups  in Geometric Algebra, and Hestenes and Holt \cite{Hestenes2002CrystGroups} have discussed the crystallographic point and space groups from a conformal point of view. Here, we are  interested in applying Geometric Algebra in the Coxeter framework, in particular in the context of  root systems, the Coxeter element,  non-crystallographic groups and quasicrystals, which to our knowledge have not yet been treated at all.

 This paper is organised as follows. 
Section \ref{sec_Cox} introduces how systems are currently modeled in terms of Coxeter groups, and what kind of computations arise in this context. 
In Section \ref{sec_versor}, we present a versor formalism in which the  full, chiral and binary polyhedral groups can all be easily generated and treated within the same framework. 
In particular, this yields a construction of the binary polyhedral groups (discrete subgroups of $SU(2)$ that are  the double covers of the chiral (rotation) groups), which we will discuss further in Section \ref{sec_bin}.
In Section \ref{sec_CGA}, we briefly outline how to extend this treatment to the conformal setup,
in particular for the non-crystallographic groups, and we demonstrate how to construct root systems and quasicrystalline point sets in this framework. 
In Section \ref{sec_2D}, we  discuss the two-dimensional non-crystallographic Coxeter groups $I_2(n)$ as well as the icosahedral group $H_3$ and the other two three-dimensional groups $A_3$ (tetrahedral) and $B_3$ (octahedral) in a versor formalism, which elucidates the relation with the Coxeter element and the Coxeter plane.
We conclude with a summary and possible further work in Section \ref{sec_concl}.

\section{Coxeter formulation}\label{sec_Cox}
Coxeter groups are abstract groups describable in terms of mirror symmetries \cite{Coxeter1934discretegroups}. 
The elements of finite Coxeter groups can be visualised as reflections at planes through the origin in a Euclidean vector space $V$. In particular, for $v$, $\alpha\in V$, then 
\begin{equation}\label{reflect}
v\rightarrow r_\alpha v = v'= v - \frac{2\alpha\cdot  v}{\alpha\cdot \alpha}\alpha
\end{equation}
corresponds to a Euclidean reflection $r_\alpha$ of the vector $v$ at a hyperplane perpendicular to the so-called root vector $\alpha$. The structure of the Coxeter group is  thus encoded in the collection  of all such roots, which form a root system. 
A subset of the root system, called the {simple roots}, is sufficient to express every root via a $\mathbb{Z}$-linear combination with coefficients of the same sign. The root system is therefore  completely characterised by this basis of simple roots, which in turn completely characterises the Coxeter group. The number of simple roots is called the rank of the root system, which essentially gives the dimension and therefore indexes the corresponding Coxeter group and root system (e.g. $H_3$ for the largest discrete symmetry group in three dimensions).

Finite Coxeter groups describe the  properties of physical structures, e.g. of a viral protein container or a carbon onion, at a given radial level, as the symmetry only relates features at the same radial distance from the origin. In order to obtain information on how  structural properties at different radial levels could collectively be constrained by symmetry, affine extensions of these groups need to be considered. 
Affine extensions are constructed in the Coxeter framework by adding affine reflection planes not containing the origin 
\cite{McCammond2010Coxeter}. A detailed account of this construction is presented elsewhere  \cite{Twarock:2002a, DechantTwarockBoehm2011H3aff, DechantTwarockBoehm2011E8A4}, but essentially the affine extension amounts to making the reflection group $G$ topologically non-compact by adding a translation operator $T$. 
The structures of viruses follow several different extensions of the (chiral) icosahedral group $I$ by  translation operators \cite{Keef:2009, DechantTwarockBoehm2011Jess, Keef:2013}. Thus, a wide range of empirical observations in virology can be explained by  affine Coxeter groups. 
We now discuss 2D counterparts to the 3D point arrays that predict the architecture of viruses and fullerenes, and explain in what sense the translation operators are distinguished.

\begin{figure}
\begin{center}
\tikzstyle{background grid}=[draw, black!50,step=.5cm]
    \begin{tabular}{@{}c@{ }c@{ }c@{ }}
		\begin{tikzpicture}[
		    knoten/.style={        circle,      inner sep=.08cm,        draw}  , 
		   dot/.style={        circle,      inner sep=.02cm,        draw}  , 
		my node/.style={trapezium, fill=#1!20, draw=#1!75, text=black} 
		   ]
		
			  \node at (0.6, 1.9)   {};
			  \node at (0.6, -1.9)   {};
		  \node at (1, 0)   (hex1)  [knoten,  ball color=black] {};
		  \node at (.309, -.951) (hex2) [knoten, ball color=black] {};  
		  \node at (.309, .951) (hex3) [knoten, ball color=black] {};
		  \node at (-.809, -.588) (hex4) [knoten,  ball color=black] {};
		  \node at (-.809, .588) (hex5) [knoten, ball color=black] {};
		  \path  (hex1) edge (hex3);
		  \path  (hex1) edge (hex2);
		  \path  (hex2) edge (hex4);
		  \path  (hex3) edge (hex5);
		  \path  (hex5) edge (hex4);
				  \path[->] (hex1) edge [bend right] node [above right] {\textcolor{red}{$G$}} (hex3) ;		
		
		\end{tikzpicture}&\hspace{0.5cm}
				\begin{tikzpicture}[
				    knoten/.style={        circle,      inner sep=.08cm,        draw}  , 
				   dot/.style={        circle,      inner sep=.02cm,        draw}  , 
				my node/.style={trapezium, fill=#1!20, draw=#1!75, text=black} 
				   ]

				  \node at (0.6, 1.9)   {};
				  \node at (0.6, -1.9)   {};
				  \node at (1, 0)   (hex1)  [knoten,  ball color=black] {};
				  \node at (.309, -.951) (hex2) [knoten, ball color=black] {};  
				  \node at (.309, .951) (hex3) [knoten, ball color=black] {};
				  \node at (-.809, -.588) (hex4) [knoten,  ball color=black] {};
				  \node at (-.809, .588) (hex5) [knoten, ball color=black] {};
				  \path  (hex1) edge (hex3);
				  \path  (hex1) edge (hex2);
				  \path  (hex2) edge (hex4);
				  \path  (hex3) edge (hex5);
				  \path  (hex5) edge (hex4);

				  \node at (1+1, 0)   (hex1e)  [knoten,  ball color=gray!30!white] {};
				  \node at (.309+1, -.951) (hex2e) [knoten, ball color=gray!30!white] {};  
				  \node at (.309+1, .951) (hex3e) [knoten, ball color=gray!30!white] {};
				  \node at (-.809+1, -.588) (hex4e) [knoten,  ball color=gray!30!white] {};
				  \node at (-.809+1, .588) (hex5e) [knoten, ball color=gray!30!white] {};
				  \path  (hex1e) edge (hex3e);
				  \path  (hex1e) edge (hex2e);
				  \path  (hex2e) edge (hex4e);
				  \path  (hex3e) edge (hex5e);
				  \path  (hex5e) edge (hex4e);

				  \path [->, color=red] (hex1) edge node [above] {\textcolor{red}{$T$}} (hex1e);		

				\end{tikzpicture} &\hspace{0.5cm}

					\begin{tikzpicture}[
					    knoten/.style={        circle,      inner sep=.08cm,        draw}  , 
					   dot/.style={        circle,      inner sep=.02cm,        draw}  , 
					my node/.style={trapezium, fill=#1!20, draw=#1!75, text=black} 
					   ]

					  \node at (1, 0)   (hex1)  [knoten,  ball color=black] {};
					  \node at (.309, -.951) (hex2) [knoten, ball color=black] {};  
					  \node at (.309, .951) (hex3) [knoten, ball color=black] {};
					  \node at (-.809, -.588) (hex4) [knoten,  ball color=black] {};
					  \node at (-.809, .588) (hex5) [knoten, ball color=black] {};
					  \path  (hex1) edge (hex3);
					  \path  (hex1) edge (hex2);
					  \path  (hex2) edge (hex4);
					  \path  (hex3) edge (hex5);
					  \path  (hex5) edge (hex4);

					  \node at (1++.309, 0-.951)   (hex1a)  [knoten,  ball color=gray!30!white] {};
					  \node at (.309+.309, -.951-.951) (hex2a) [knoten, ball color=gray!30!white] {};  
					  \node at (.309+.309, .951-.951) (hex3a) [knoten, ball color=gray!30!white] {};
					  \node at (-.809+.309, -.588-.951) (hex4a) [knoten,  ball color=gray!30!white] {};
					  \node at (-.809+.309, .588-.951) (hex5a) [knoten, ball color=gray!30!white] {};
					  \path  (hex1a) edge (hex3a);
					  \path  (hex1a) edge (hex2a);
					  \path  (hex2a) edge (hex4a);
					  \path  (hex3a) edge (hex5a);
					  \path  (hex5a) edge (hex4a);

					  \node at (1+1, 0)   (hex1e)  [knoten,  ball color=gray!30!white] {};
					  \node at (.309+1, -.951) (hex2e) [knoten, ball color=gray!30!white] {};  
					  \node at (.309+1, .951) (hex3e) [knoten, ball color=gray!30!white] {};
					  \node at (-.809+1, -.588) (hex4e) [knoten,  ball color=gray!30!white] {};
					  \node at (-.809+1, .588) (hex5e) [knoten, ball color=gray!30!white] {};
					  \path  (hex1e) edge (hex3e);
					  \path  (hex1e) edge (hex2e);
					  \path  (hex2e) edge (hex4e);
					  \path  (hex3e) edge (hex5e);
					  \path  (hex5e) edge (hex4e);

					  \node at (1+.309, 0+.951)   (hex1b)  [knoten,  ball color=gray!30!white] {};
					  \node at (.309+.309, -.951+.951) (hex2b) [knoten, ball color=gray!30!white] {};  
					  \node at (.309+.309, .951+.951) (hex3b) [knoten, ball color=gray!30!white] {};
					  \node at (-.809+.309, -.588+.951) (hex4b) [knoten,  ball color=gray!30!white] {};
					  \node at (-.809+.309, .588+.951) (hex5b) [knoten, ball color=gray!30!white] {};

					  \path  (hex1b) edge (hex3b);
					  \path  (hex1b) edge (hex2b);
					  \path  (hex2b) edge (hex4b);
					  \path  (hex3b) edge (hex5b);
					  \path  (hex5b) edge (hex4b);

					  \node at (1-.809, 0-.588)   (hex1c)  [knoten,  ball color=gray!30!white] {};
					  \node at (.309-.809, -.951-.588) (hex2c) [knoten, ball color=gray!30!white] {};  
					  \node at (.309-.809, .951-.588) (hex3c) [knoten, ball color=gray!30!white] {};
					  \node at (-.809-.809, -.588-.588) (hex4c) [knoten,  ball color=gray!30!white] {};
					  \node at (-.809-.809, .588-.588) (hex5c) [knoten, ball color=gray!30!white] {};
					  \path  (hex1c) edge (hex3c);
					  \path  (hex1c) edge (hex2c);
					  \path  (hex2c) edge (hex4c);
					  \path  (hex3c) edge (hex5c);
					  \path  (hex5c) edge (hex4c);

					  \node at (1-.809, 0+.588)   (hex1d)  [knoten,  ball color=gray!30!white] {};
					  \node at (.309-.809, -.951+.588) (hex2d) [knoten, ball color=gray!30!white] {};  
					  \node at (.309-.809, .951+.588) (hex3d) [knoten, ball color=gray!30!white] {};
					  \node at (-.809-.809, -.588+.588) (hex4d) [knoten,  ball color=gray!30!white] {};
					  \node at (-.809-.809, .588+.588) (hex5d) [knoten, ball color=gray!30!white] {};
					  \path  (hex1d) edge (hex3d);
					  \path  (hex1d) edge (hex2d);
					  \path  (hex2d) edge (hex4d);
					  \path  (hex3d) edge (hex5d);
					  \path  (hex5d) edge (hex4d);
					  \path[->] (hex1e) edge [bend right] node [above right] {\textcolor{red}{$G$}} (hex3b) ;

					\end{tikzpicture}
		\\

\end{tabular}	

\end{center}

\caption[dummy1]{The action of an affine Coxeter group on a pentagon. The translation operator $T$ generates extended point arrays, whilst the compact part $G$ makes the resulting point set rotationally symmetric. Blueprints with degeneracies due to coinciding points correspond to non-trivial group structures and can be used in the modeling of viruses.}
\label{figpent}
\end{figure}
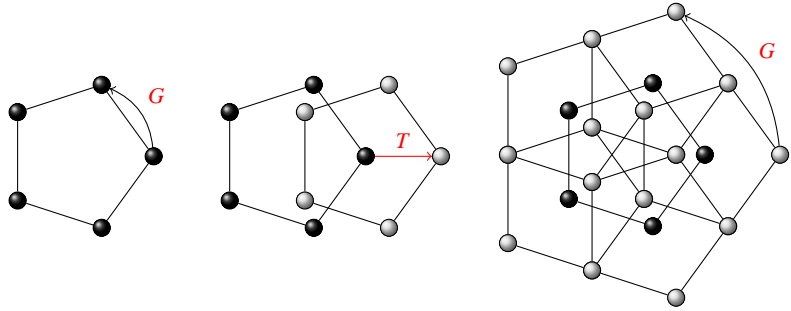

For illustration purposes, let us consider a similar construction for a  pentagon of unit size, as shown in Fig. \ref{figpent}. 
The non-compact translation operator $T$, here chosen to also be of unit length, creates a displaced version of the pentagon. 
The action of the  symmetry group $G$ of the pentagon then generates further copies in such a way that the final point array displays the same rotational symmetries.

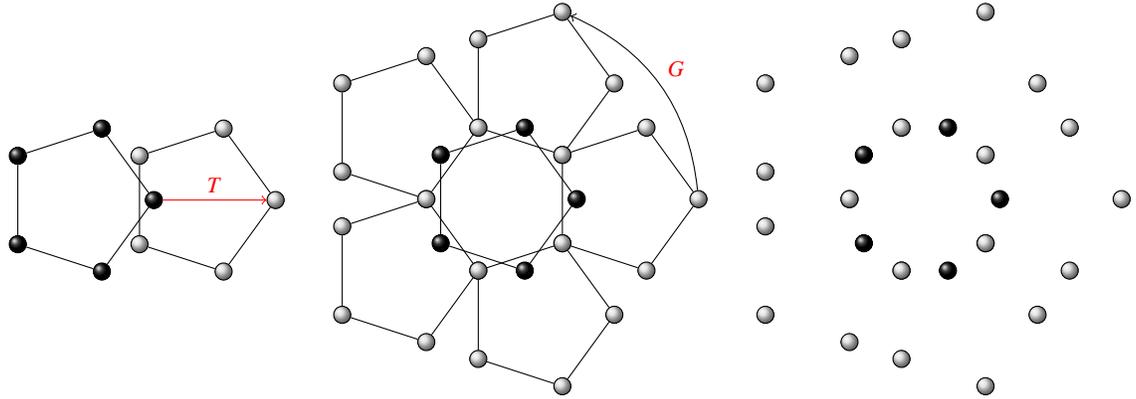
\begin{figure}
\begin{center}
\tikzstyle{background grid}=[draw, black!50,step=.5cm]
    \begin{tabular}{@{}c@{ }c@{ }c@{ }}
			\begin{tikzpicture}[
			    knoten/.style={        circle,      inner sep=.08cm,        draw}  , 
			   dot/.style={        circle,      inner sep=.02cm,        draw}  , 
			my node/.style={trapezium, fill=#1!20, draw=#1!75, text=black} 
			   ]

			  \node at (0.8, 2.5)   {};
			  \node at (0.8, -2.5)   {};
			  \node at (1, 0)   (hex1)  [knoten,  ball color=black] {};
			  \node at (.309, -.951) (hex2) [knoten, ball color=black] {};  
			  \node at (.309, .951) (hex3) [knoten, ball color=black] {};
			  \node at (-.809, -.588) (hex4) [knoten,  ball color=black] {};
			  \node at (-.809, .588) (hex5) [knoten, ball color=black] {};
			  \path  (hex1) edge (hex3);
			  \path  (hex1) edge (hex2);
			  \path  (hex2) edge (hex4);
			  \path  (hex3) edge (hex5);
			  \path  (hex5) edge (hex4);

			  \node at (1+1.618, 0)   (hex1e)  [knoten,  ball color=gray!30!white] {};
			  \node at (.309+1.618, -.951) (hex2e) [knoten, ball color=gray!30!white] {};  
			  \node at (.309+1.618, .951) (hex3e) [knoten, ball color=gray!30!white] {};
			  \node at (-.809+1.618, -.588) (hex4e) [knoten,  ball color=gray!30!white] {};
			  \node at (-.809+1.618, .588) (hex5e) [knoten, ball color=gray!30!white] {};
			  \path  (hex1e) edge (hex3e);
			  \path  (hex1e) edge (hex2e);
			  \path  (hex2e) edge (hex4e);
			  \path  (hex3e) edge (hex5e);
			  \path  (hex5e) edge (hex4e);

			  \path [->, color=red] (hex1) edge node [above] {\textcolor{red}{$T$}} (hex1e);		
			\end{tikzpicture}
		&\hspace{0.5cm}
					\begin{tikzpicture}[
					    knoten/.style={        circle,      inner sep=.08cm,        draw}  , 
					   dot/.style={        circle,      inner sep=.02cm,        draw}  , 
					my node/.style={trapezium, fill=#1!20, draw=#1!75, text=black} 
					   ]

					  \node at (1, 0)   (hex1)  [knoten,  ball color=black] {};
					  \node at (.309, -.951) (hex2) [knoten, ball color=black] {};  
					  \node at (.309, .951) (hex3) [knoten, ball color=black] {};
					  \node at (-.809, -.588) (hex4) [knoten,  ball color=black] {};
					  \node at (-.809, .588) (hex5) [knoten, ball color=black] {};
					  \path  (hex1) edge (hex3);
					  \path  (hex1) edge (hex2);
					  \path  (hex2) edge (hex4);
					  \path  (hex3) edge (hex5);
					  \path  (hex5) edge (hex4);

					  \node at (1+1.618*.309, 0-1.618*.951)   (hex1a)  [knoten,  ball color=gray!30!white] {};
					  \node at (.309+1.618*.309, -.951-1.618*.951) (hex2a) [knoten, ball color=gray!30!white] {};  
					  \node at (.309+1.618*.309, .951-1.618*.951) (hex3a) [knoten, ball color=gray!30!white] {};
					  \node at (-.809+1.618*.309, -.588-1.618*.951) (hex4a) [knoten,  ball color=gray!30!white] {};
					  \node at (-.809+1.618*.309, .588-1.618*.951) (hex5a) [knoten, ball color=gray!30!white] {};
					  \path  (hex1a) edge (hex3a);
					  \path  (hex1a) edge (hex2a);
					  \path  (hex2a) edge (hex4a);
					  \path  (hex3a) edge (hex5a);
					  \path  (hex5a) edge (hex4a);

					  \node at (1+1.618*1, 0)   (hex1e)  [knoten,  ball color=gray!30!white] {};
					  \node at (.309+1.618*1, -.951) (hex2e) [knoten, ball color=gray!30!white] {};  
					  \node at (.309+1.618*1, .951) (hex3e) [knoten, ball color=gray!30!white] {};
					  \node at (-.809+1.618*1, -.588) (hex4e) [knoten,  ball color=gray!30!white] {};
					  \node at (-.809+1.618*1, .588) (hex5e) [knoten, ball color=gray!30!white] {};
					  \path  (hex1e) edge (hex3e);
					  \path  (hex1e) edge (hex2e);
					  \path  (hex2e) edge (hex4e);
					  \path  (hex3e) edge (hex5e);
					  \path  (hex5e) edge (hex4e);

					  \node at (1+1.618*.309, 0+1.618*.951)   (hex1b)  [knoten,  ball color=gray!30!white] {};
					  \node at (.309+1.618*.309, -.951+1.618*.951) (hex2b) [knoten, ball color=gray!30!white] {};  
					  \node at (.309+1.618*.309, .951+1.618*.951) (hex3b) [knoten, ball color=gray!30!white] {};
					  \node at (-.809+1.618*.309, -.588+1.618*.951) (hex4b) [knoten,  ball color=gray!30!white] {};
					  \node at (-.809+1.618*.309, .588+1.618*.951) (hex5b) [knoten, ball color=gray!30!white] {};

					  \path  (hex1b) edge (hex3b);
					  \path  (hex1b) edge (hex2b);
					  \path  (hex2b) edge (hex4b);
					  \path  (hex3b) edge (hex5b);
					  \path  (hex5b) edge (hex4b);

					  \node at (1-1.618*.809, 0-1.618*.588)   (hex1c)  [knoten,  ball color=gray!30!white] {};
					  \node at (.309-1.618*.809, -.951-1.618*.588) (hex2c) [knoten, ball color=gray!30!white] {};  
					  \node at (.309-1.618*.809, .951-1.618*.588) (hex3c) [knoten, ball color=gray!30!white] {};
					  \node at (-.809-1.618*.809, -.588-1.618*.588) (hex4c) [knoten,  ball color=gray!30!white] {};
					  \node at (-.809-1.618*.809, .588-1.618*.588) (hex5c) [knoten, ball color=gray!30!white] {};
					  \path  (hex1c) edge (hex3c);
					  \path  (hex1c) edge (hex2c);
					  \path  (hex2c) edge (hex4c);
					  \path  (hex3c) edge (hex5c);
					  \path  (hex5c) edge (hex4c);

					  \node at (1-1.618*.809, 0+1.618*.588)   (hex1d)  [knoten,  ball color=gray!30!white] {};
					  \node at (.309-1.618*.809, -.951+1.618*.588) (hex2d) [knoten, ball color=gray!30!white] {};  
					  \node at (.309-1.618*.809, .951+1.618*.588) (hex3d) [knoten, ball color=gray!30!white] {};
					  \node at (-.809-1.618*.809, -.588+1.618*.588) (hex4d) [knoten,  ball color=gray!30!white] {};
					  \node at (-.809-1.618*.809, .588+1.618*.588) (hex5d) [knoten, ball color=gray!30!white] {};
					  \path  (hex1d) edge (hex3d);
					  \path  (hex1d) edge (hex2d);
					  \path  (hex2d) edge (hex4d);
					  \path  (hex3d) edge (hex5d);
					  \path  (hex5d) edge (hex4d);
					  \path[->] (hex1e) edge [bend right] node [above right] {\textcolor{red}{$G$}} (hex3b) ;		

					\end{tikzpicture} &\hspace{0.5cm}

					\begin{tikzpicture}[
					    knoten/.style={        circle,      inner sep=.08cm,        draw}  , 
					   dot/.style={        circle,      inner sep=.02cm,        draw}  , 
					my node/.style={trapezium, fill=#1!20, draw=#1!75, text=black} 
					   ]

					  \node at (1, 0)   (hex1)  [knoten,  ball color=black] {};
					  \node at (.309, -.951) (hex2) [knoten, ball color=black] {};  
					  \node at (.309, .951) (hex3) [knoten, ball color=black] {};
					  \node at (-.809, -.588) (hex4) [knoten,  ball color=black] {};
					  \node at (-.809, .588) (hex5) [knoten, ball color=black] {};

					  \node at (1+1.618*.309, 0-1.618*.951)   (hex1a)  [knoten,  ball color=gray!30!white] {};
					  \node at (.309+1.618*.309, -.951-1.618*.951) (hex2a) [knoten, ball color=gray!30!white] {};  
					  \node at (.309+1.618*.309, .951-1.618*.951) (hex3a) [knoten, ball color=gray!30!white] {};
					  \node at (-.809+1.618*.309, -.588-1.618*.951) (hex4a) [knoten,  ball color=gray!30!white] {};
					  \node at (-.809+1.618*.309, .588-1.618*.951) (hex5a) [knoten, ball color=gray!30!white] {};

					  \node at (1+1.618*1, 0)   (hex1e)  [knoten,  ball color=gray!30!white] {};
					  \node at (.309+1.618*1, -.951) (hex2e) [knoten, ball color=gray!30!white] {};  
					  \node at (.309+1.618*1, .951) (hex3e) [knoten, ball color=gray!30!white] {};
					  \node at (-.809+1.618*1, -.588) (hex4e) [knoten,  ball color=gray!30!white] {};
					  \node at (-.809+1.618*1, .588) (hex5e) [knoten, ball color=gray!30!white] {};

					  \node at (1+1.618*.309, 0+1.618*.951)   (hex1b)  [knoten,  ball color=gray!30!white] {};
					  \node at (.309+1.618*.309, -.951+1.618*.951) (hex2b) [knoten, ball color=gray!30!white] {};  
					  \node at (.309+1.618*.309, .951+1.618*.951) (hex3b) [knoten, ball color=gray!30!white] {};
					  \node at (-.809+1.618*.309, -.588+1.618*.951) (hex4b) [knoten,  ball color=gray!30!white] {};
					  \node at (-.809+1.618*.309, .588+1.618*.951) (hex5b) [knoten, ball color=gray!30!white] {};

					  \node at (1-1.618*.809, 0-1.618*.588)   (hex1c)  [knoten,  ball color=gray!30!white] {};
					  \node at (.309-1.618*.809, -.951-1.618*.588) (hex2c) [knoten, ball color=gray!30!white] {};  
					  \node at (.309-1.618*.809, .951-1.618*.588) (hex3c) [knoten, ball color=gray!30!white] {};
					  \node at (-.809-1.618*.809, -.588-1.618*.588) (hex4c) [knoten,  ball color=gray!30!white] {};
					  \node at (-.809-1.618*.809, .588-1.618*.588) (hex5c) [knoten, ball color=gray!30!white] {};

					  \node at (1-1.618*.809, 0+1.618*.588)   (hex1d)  [knoten,  ball color=gray!30!white] {};
					  \node at (.309-1.618*.809, -.951+1.618*.588) (hex2d) [knoten, ball color=gray!30!white] {};  
					  \node at (.309-1.618*.809, .951+1.618*.588) (hex3d) [knoten, ball color=gray!30!white] {};
					  \node at (-.809-1.618*.809, -.588+1.618*.588) (hex4d) [knoten,  ball color=gray!30!white] {};
					  \node at (-.809-1.618*.809, .588+1.618*.588) (hex5d) [knoten, ball color=gray!30!white] {};

					\end{tikzpicture}
					
		\\

\end{tabular}	

\end{center}

\caption[dummy1]{Translation by the golden ratio results in a point set whose constituent polygons are simultaneously constrained by the affine symmetry.}
\label{figpent2}
\end{figure}

The translation operator we have chosen for this example is distinguished  because several of the generated points lie on more than one pentagon, for instance the innermost points, or the midpoints of the edges of the large outer pentagon.  
 Certain distinguished translations lead to such point sets with  degenerate points, which therefore have lower cardinality than those obtained by a random translation (here 15 points as opposed to 25).
This degeneracy yields a non-trivial mathematical structure at the group level, and  the corresponding blueprints in three dimensions can be used to model icosahedral viruses. 

Fig. \ref{figpent2} shows a similar example for a translation of length of the golden ratio $\tau=\frac{1}{2}(1+\sqrt{5})\approx 1.618$. 
The resulting point set also has degenerate cardinality (now 20 points), and consists of an inner decagon and an outer pentagon. 
Affine symmetry here means that the relative sizes of the decagon and pentagon are fixed by the group structure. 
This is a powerful geometric tool for constraining real systems.

The computations necessary in this context are therefore translations, reflections and rotations; one also needs to be able to check degeneracy of points. 
In the usual vector space approach, these operations are implemented via matrices. 
We instead develop here a versor implementation. 
This has some computational advantages, as well as offering surprising geometric insights, as we shall see later.
Whilst the computational complexity for 3-dimensional applications is limited, equivalent computations in four dimensions, where $H_4$ -- the four-dimensional analogue of the icosahedral group and symmetry group of the hypericosahedron (600-cell) -- has order $14,400$ and $H_4$-symmetric polytopes have upwards of $120$ and $600$ vertices, are rather more complex. 

In the Coxeter setting, therefore, the reflections are fundamental; Geometric Algebra is very efficient at encoding reflections algebraically, and at performing computations with clear geometric content. However, the two frameworks do not appear to have been combined previously. We therefore explore which benefits a Clifford algebraic description might offer for Coxeter group theoretic considerations.

\section{Versor framework}\label{sec_versor}
The geometric product $xy=x\cdot y+x \wedge y$ of two vectors $x$ and $y$ (with $x\cdot y$ denoting the scalar product and $x \wedge y$ the exterior product) of Geometric Algebra \cite{Hestenes1966STA, HestenesSobczyk1984, Hestenes1990NewFound,LasenbyDoran2003GeometricAlgebra}
 provides a very compact and efficient way of handling reflections in any number of dimensions, and thus by the Cartan-Dieudonn\'e theorem also rotations \cite{Garling2011Clifford}. For a unit vector $\alpha$, the two terms in the formula for a reflection of a vector $v$ in the hyperplane orthogonal to $\alpha$ from Eq. (\ref{reflect}) simplify to the double-sided action of $\alpha$ via the geometric product
	\begin{equation}\label{in2refl}
	  v\rightarrow r_\alpha v=v'=-\alpha v \alpha.
	\end{equation}
This  prescription for reflecting vectors in hyperplanes is remarkably compact, and applies more generally to all multivectors.  Even more importantly, from the  Cartan-Dieudonn\'e theorem, rotations are the product of an even number of successive reflections. For instance, compounding the reflections in the hyperplanes defined by the unit vectors $\alpha_i$ and $\alpha_j$ results in a rotation in the plane defined by $\alpha_i\wedge \alpha_j$
	\begin{equation}\label{in2rot}
	  v''=\alpha_j\alpha_iv \alpha_i\alpha_j=: \tilde{R}v{R},
	\end{equation}
where we have defined the rotor $R=\alpha_i\alpha_j$ and the tilde denotes the reversal of the order of the constituent vectors of a versor, e.g. here $\tilde{R}=\alpha_j\alpha_i$.
Rotors satisfy $\tilde{R}R=R\tilde{R}=1$ and themselves transform single-sidedly under further rotations. They thus form a multiplicative group under the geometric product, called the rotor group, which is essentially the Spin group, and thus a double-cover of the special orthogonal group \cite{Hestenes1966STA, LasenbyDoran2003GeometricAlgebra, Porteous1995Clifford}. Objects in Geometric Algebra that transform single-sidedly are called spinors, so that rotors are  normalised spinors. 

In fact, the above two cases are examples of a more general theorem on the Geometric Algebra representation of orthogonal transformations. 
In analogy to the vectors and rotors above, a versor is a multivector $A=a_1a_2\dots a_k$ which is the product of $k$ non-null vectors $a_i$ ($a_i^2\ne 0$). These versors also form a multiplicative group under the geometric product, called the versor group. 
The Versor Theorem \cite{Hestenes1990NewFound} then states that
every orthogonal transformation $\underbar{A}$ of a vector $v$ can be expressed via unit versors in the canonical form
\begin{equation}\label{in2versor}
\underbar{A}: v\rightarrow  v'=\underbar{A}(v)=\pm{\tilde{A}vA},
\end{equation}
where the $\pm$-sign defines the parity of the versor.
Since both the versors $A$ and $-A$ encode the same orthogonal transformation $\underbar{A}$, unit versors are double-valued representations of the respective orthogonal transformation, giving a construction of the Pin group $\Pin(p,q)$ \cite{Porteous1995Clifford}, the double cover of the orthogonal group $O(p,q)$. Even versors form a double covering of the special orthogonal group $SO(p,q)$, called the Spin group $\Spin(p,q)$.
The versor realisation of the orthogonal group is much simpler than conventional matrix approaches. This is  particularly useful in the Conformal Geometric Algebra setup, where one uses the fact that the conformal group   $C(p,q)$ is homomorphic to $O(p+1,q+1)$ to treat translations as well as rotations in a unified versor framework (see Section \ref{sec_CGA}), making it possible to use all of GA's versor machinery for the analysis of the conformal group.

We now consider which benefits such a versor approach can offer for Coxeter computations, in particular in the context of applications to physical phenomena in three dimensions. The isometry group of three-dimensional space is the orthogonal group $O(3)$, of which the full polyhedral (Coxeter) groups  are discrete subgroups. 
However, $O(3)$ is globally $SO(3)\times \mathbb{Z}_2$, where the special orthogonal group $SO(3)$ is the subgroup of pure rotations (or the chiral part). 
$SO(3)$ is still not simply-connected, but is  doubly covered by the Spin group $\Spin(3)\simeq SU(2)$ (in fact, it is $SO(3)\times \mathbb{Z}_2$ locally, i.e. a fibre bundle). 
Thus, the chiral polyhedral groups are discrete subgroups of $SO(3)$, the full polyhedral groups (Coxeter) are their preimage in $O(3)$, and the binary polyhedral groups are their preimage under the universal covering in $\Spin(3)$.

\begin{table}
\caption{Versor framework for a unified treatment of the chiral, full and binary polyhedral groups.}
\label{tab:1}       
%
%
\begin{tabular}{p{1.5cm}p{3cm}p{6.9cm}}
\hline\noalign{\smallskip}
Group &Discrete subgroup & Action Mechanism  \\
\noalign{\smallskip}\svhline\noalign{\smallskip}
$SO(3)$&rotational (chiral) & $x\rightarrow \tilde{R}xR$\\
$O(3)$&reflection (full) & $x\rightarrow \pm\tilde{A}xA$\\
$\Spin(3)$&binary  & spinors $R$ under spinor multiplication $(R_1,R_2)\rightarrow R_1R_2$\\
\noalign{\smallskip}\hline\noalign{\smallskip}
\end{tabular}
\end{table}

We begin with the simple roots (vertex vectors) which completely characterise a given Coxeter group, and consider their closure under mutual reflections (the root system). We then compute the rotors derivable from all these root vectors/reflections, which encode the rotational part of the respective polyhedral group via the double-sided action in Eq. (\ref{in2rot}). The rotor group defined by single-sided action can in fact be shown to realise the respective binary polyhedral group, which is the double cover of the chiral polyhedral group under the universal covering homomorphism between $SO(3)$ and $\Spin(3)$. Finally, including the versors of the form $\alpha_i\alpha_j\alpha_k$  via double-sided action gives a realisation of the full polyhedral group (the Coxeter group).  The proofs are straightforward calculations in the Geometric Algebra of three dimensions and more details are contained in \cite{Dechant2012CoxGA, Dechant2013Polytopes}.

\begin{thm}[Reflections/Coxeter groups and polyhedra/root systems]
Take the three simple roots for the Coxeter groups $A_1\times A_1\times A_1$ (respectively $A_3$/$B_3$/$H_3$). Geometric Algebra reflections in the hyperplanes orthogonal to these vectors via Eq. (\ref{in2refl}) generate further vectors pointing to the 6 (resp. 12/18/30) vertices of an octahedron (resp. cuboctahedron/cuboctahedron with an  octahedron/icosidodecahedron), giving the full root system of the group.
\end{thm}
For instance, the simple roots for $A_1\times A_1\times A_1$ are $\alpha_1=e_1$, $\alpha_2=e_2$ and $\alpha_3=e_3$ for orthonormal basis vectors $e_i$. Reflections amongst those then also generate $-e_1$, $-e_2$ and $-e_3$, which all together point to the vertices of an octahedron.

By the Cartan-Dieudonn\'e theorem, combining two reflections yields a rotation, and Eq. (\ref{in2rot}) gives a rotor realisation of these rotations in Geometric Algebra.
\begin{thm}[Spinors from reflections]\label{HGA_rotors}
The 6 (resp. 12/18/30)  reflections in the  Coxeter group  $A_1\times A_1 \times A_1$ (resp. $A_3$/$B_3$/$H_3$)   generate 8 (resp. 24/48/120) rotors.
\end{thm}
For the $A_1\times A_1\times A_1$ example above, the spinors thus generated are $\pm 1$, $\pm e_1e_2$, $\pm e_2 e_3$ and $\pm e_3e_1$. In fact, these groups of discrete spinors yield a novel construction of the binary polyhedral groups.

\begin{thm}[Spinor groups and binary polyhedral groups]
The discrete spinor group in Theorem \ref{HGA_rotors} is isomorphic to the quaternion group $Q$ (resp. binary tetrahedral group $2T$/binary octahedral group $2O$/binary icosahedral group $2I$).
\end{thm}

Through the versor theorem, we can therefore describe all three types of groups in the same framework. 
Vectors are grade 1 versors, and rotors are grade 2 versors. For instance, the 60 rotations  of the chiral icosahedral group $I$  are given by 120 rotors acting as $\alpha_i\alpha_j v \alpha_j \alpha_i$.  60 operations of odd parity are defined by 120 grade 1 and grade 3 versors (with vector and trivector parts) acting as $-\alpha_i\alpha_j\alpha_k v \alpha_k \alpha_j \alpha_i$. However, 30 of them are just the 15 true reflections given by pure vectors, leaving another 45 rotoinversions. Thus, the Coxeter group (the full icosahedral group $H_3\subset O(3)$) is expressed in accordance with the versor theorem. Alternatively, one can think of 60 rotations and 60 rotoinversions, making  $H_3=I_h=I\times\mathbb{Z}_2$ manifest. However, the rotations  operate double-sidedly  on a vector, such that the versor formalism actually provides a 2-valued representation of the rotation group $SO(3)$, since the rotors $R$ and $-R$ encode the same rotation. Since $\Spin(3)$ is  the universal 2-cover of $SO(3)$, the rotors form a realisation of the preimage of the chiral icosahedral group $I$, i.e. the binary icosahedral group $2I$. Thus, in the versor approach, we can treat all these different groups in a unified framework, whilst maintaining a clear conceptual separation. In Table \ref{tab:1}, we summarise how the three different types of polyhedral groups are realised in the versor framework.

\section{Construction of the binary polyhedral groups}\label{sec_bin}
In this section, we consider further the implications of our  construction of the binary polyhedral groups.
Since Clifford algebra is well known to provide a simple construction of the Spin groups, it is perhaps not surprising -- from a Clifford algebra point of view -- to find that the discrete rotor groups realise the binary polyhedral groups.
However, this construction does not seem to be known, and from a Coxeter group point of view, it leads to rather surprising consequences.

\begin{table}
\begin{centering}\begin{tabular}{|c|c||c||c|c|c|}
\hline
rank-3 group&diagram&binary&rank-4 group&diagram
\tabularnewline
\hline
\hline
$A_1\times A_1\times A_1$&\begin{tikzpicture}[
    knoten/.style={        circle,      inner sep=.05cm,        draw}  
   ]
  \node at (1,1.5) (knoten1) [knoten,  color=white!0!black] {};
  \node at (1.5,1.5) (knoten2) [knoten,  color=white!0!black] {};
  \node at (2,1.5) (knoten3) [knoten,  color=white!0!black] {};
\end{tikzpicture}&$Q$&$A_1\times A_1 \times A_1\times A_1$&\begin{tikzpicture}[
    knoten/.style={        circle,      inner sep=.05cm,        draw}  
   ]
  \node at (1,1.5) (knoten1) [knoten,  color=white!0!black] {};
  \node at (1.5,1.5) (knoten2) [knoten,  color=white!0!black] {};
  \node at (2,1.5) (knoten3) [knoten,  color=white!0!black] {};
  \node at (2.5,1.5) (knoten4) [knoten,  color=white!0!black] {};
\end{tikzpicture}
\tabularnewline
\hline
$A_3$&\begin{tikzpicture}[
    knoten/.style={        circle,      inner sep=.05cm,        draw}  
   ]
  \node at (1,1.5) (knoten1) [knoten,  color=white!0!black] {};
  \node at (1.5,1.5) (knoten2) [knoten,  color=white!0!black] {};
  \node at (2,1.5) (knoten3) [knoten,  color=white!0!black] {};
		  \path  (knoten1) edge (knoten2);
		  \path  (knoten2) edge (knoten3);

\end{tikzpicture}&$2T$&$D_4$&		\begin{tikzpicture}[
		    knoten/.style={        circle,      inner sep=.05cm,        draw}  
		   ]
		  \node at (1,1.5) (knoten1) [knoten,  color=white!0!black] {};
		  \node at (1.5,1.5) (knoten2) [knoten,  color=white!0!black] {};
		  \node at (2,1.7) (knoten3) [knoten,  color=white!0!black] {};
				  \node at (2,1.85) (knoten5)  {};
		  \node at (2,1.3) (knoten4) [knoten,  color=white!0!black] {};
			  \path  (knoten1) edge (knoten2);
			  \path  (knoten2) edge (knoten3);
				  \path  (knoten4) edge (knoten2);
				  \path  (knoten2) edge (knoten3);		
		
		\end{tikzpicture}
\tabularnewline
\hline
$B_3$&\begin{tikzpicture}[
    knoten/.style={        circle,      inner sep=.05cm,        draw}  
   ]
  \node at (1,1.5) (knoten1) [knoten,  color=white!0!black] {};
  \node at (1.5,1.5) (knoten2) [knoten,  color=white!0!black] {};
  \node at (2,1.5) (knoten3) [knoten,  color=white!0!black] {};
		  \path  (knoten1) edge (knoten2);
		  \path  (knoten2) edge node [above] {$4$} (knoten3);
\end{tikzpicture}
&$2O$&$F_4$&
\begin{tikzpicture}[
    knoten/.style={        circle,      inner sep=.05cm,        draw}  
   ]
  \node at (1,1.5) (knoten1) [knoten,  color=white!0!black] {};
  \node at (1.5,1.5) (knoten2) [knoten,  color=white!0!black] {};
  \node at (2,1.5) (knoten3) [knoten,  color=white!0!black] {};
  \node at (2.5,1.5) (knoten4) [knoten,  color=white!0!black] {};
  \path  (knoten1) edge (knoten2);
  \path  (knoten2) edge node [above] {$4$} (knoten3);
  \path  (knoten3) edge (knoten4);

\end{tikzpicture}
\tabularnewline
\hline
$H_3$&\begin{tikzpicture}[
    knoten/.style={        circle,      inner sep=.05cm,        draw}  
   ]
  \node at (1,1.5) (knoten1) [knoten,  color=white!0!black] {};
  \node at (1.5,1.5) (knoten2) [knoten,  color=white!0!black] {};
  \node at (2,1.5) (knoten3) [knoten,  color=white!0!black] {};
		  \path  (knoten1) edge (knoten2);
		  \path  (knoten2) edge node [above] {$5$} (knoten3);
\end{tikzpicture}&$2I$&$H_4$&		\begin{tikzpicture}[
		    knoten/.style={        circle,      inner sep=.05cm,        draw}  
		   ]
		  \node at (1,1.5) (knoten1) [knoten,  color=white!0!black] {};
		  \node at (1.5,1.5) (knoten2) [knoten,  color=white!0!black] {};
		  \node at (2,1.5) (knoten3) [knoten,  color=white!0!black] {};
		  \node at (2.5,1.5) (knoten4) [knoten,  color=white!0!black] {};
		  \path  (knoten1) edge (knoten2);
		  \path  (knoten4) edge node [above] {$5$} (knoten3);
		  \path  (knoten3) edge (knoten2);

		\end{tikzpicture}
\tabularnewline
\hline
\end{tabular}\par\end{centering}
\caption[Correspondence]{\label{tab_Corr} Correspondence between the rank-3 and rank-4 Coxeter groups. The spinors generated from the reflections contained in the respective rank-3 Coxeter group via the geometric product are realisations of the binary polyhedral groups $Q$, $2T$, $2O$ and $2I$, which in turn generate (mostly exceptional) rank-4 groups.}
\end{table}

The Geometric Algebra construction of the binary polyhedral groups is via rotors with (single-sided) rotor multiplication.
It is then straightforward to check the group axioms, multiplication table, conjugacy classes and the representation theory. However, it is also known that the binary polyhedral groups generate some Coxeter groups of rank 4,
for instance via quaternionic root systems \cite{Dechant2012CoxGA}. 
In particular $Q$, $2T$, $2O$ and $2I$ generate $A_1\times A_1 \times A_1\times A_1$, $D_4$, $F_4$ and $H_4$, respectively, as summarised in Table \ref{tab_Corr}. From a Coxeter perspective, this is surprising. 
However, in Geometric Algebra, spinors $\psi$ have a natural 4-dimensional Euclidean structure given by $\psi\tilde{\psi}$, and can thus also be interpreted as vectors in a 4D Euclidean space. In fact, one can show that these vertex vectors are again root systems \cite{Dechant2012Induction, Dechant2013Polytopes, Humphreys1990Coxeter}, which generate the respective rank-4 Coxeter groups. 
This demonstrates how in fact the rank-4 groups can be derived from the rank-3 groups via the geometric product of Clifford's Geometric Algebra.
This connection has so far been overlooked in  Coxeter theory.  This `induction' of higher-dimensional root systems via spinors of lower-dimensional root systems is complementary to the well-known top-down approaches of projection (for instance from $E_8$ to $H_4$ \cite{Shcherbak:1988,MoodyPatera:1993b, Koca:1998, Koca:2001, DechantTwarockBoehm2011E8A4}), or of taking subgroups by deleting nodes in Coxeter-Dynkin diagrams.
It is particularly interesting that this inductive construction relates the exceptional low-dimensional Coxeter groups $H_3$, $D_4$, $F_4$ and $H_4$ to each other as well as to the  series $A_n$, $B_n$ and $D_n$ in novel ways. In particular, it is remarkable that the exceptional  dimension-four phenomena $D_4$ (triality), $F_4$ (the largest crystallographic Coxeter group in 4D) and $H_4$ (the largest non-crystallographic Coxeter group) are seen to arise from three-dimensional geometric considerations alone, and it is possible that their existence is due to the `accidentalness' of the spinor construction. This spinorial view could thus open up novel applications in Coxeter and Lie group theory, as well as in polytopes (e.g. $A_4$), string theory and triality ($D_4$), lattice theory ($F_4$) and quasicrystals ($H_4$). 
In particular, this spinorial construction explains the symmetries of these root systems, which otherwise appear rather mysterious \cite{Dechant2013Polytopes}. 
The  $I_2(n)$ are  self-dual under the corresponding two-dimensional spinor construction \cite{Dechant2012Induction}. 

\section{Conformal Geometric Algebra and Coxeter groups}\label{sec_CGA}
The versor formalism is particularly powerful in the Conformal Geometric Algebra approach \cite{HestenesSobczyk1984, LasenbyDoran2003GeometricAlgebra,Dechant2011Thesis}. The conformal group $C(p,q)$ is $1-2$-homomorphic to $O(p+1, q+1)$ \cite{Angles1980conf, angles2008conformalbook}, for which one can easily construct the Clifford algebra and find rotor implementations of the conformal group action, including rotations and translations. Thus,  translations can also be handled multiplicatively as rotors, for flat, spherical and hyperbolic space-times, making available the `sandwiching machinery' of GA and simplifying considerably more traditional approaches and allowing novel geometric insight. Hestenes \cite{Hestenes2002PointGroups,Hestenes2002CrystGroups} has applied this framework to point and space groups, which is fruitful for the crystallographic groups, as lattice translations can be treated on the same footing as the rotations and reflections, and this approach has helped visualise these space groups \cite{Hitzer2010CLUCalc}. 

However, the non-crystallographic groups and the root system/Coxeter framework have thus far been neglected in the conformal setup. We have argued earlier and in the papers 
\cite{Twarock:2002a, DechantTwarockBoehm2011H3aff, DechantTwarockBoehm2011E8A4,Keef:2009, DechantTwarockBoehm2011Jess, Keef:2013} that in the affine extension framework translations are interesting even for the non-crystallographic groups, leading to quasicrystal-like point arrays that give blueprints for viruses and other three-dimensional physical phenomena. 
An extension of the conformal framework to translations in the case of non-crystallographic Coxeter groups could therefore have interesting consequences, including for quasilattice theory \cite{Katz:1989, Senechal:1996}, in particular when  quasicrystals are induced  via projection from higher dimensions (e.g. via the  cut-and-project method) \cite{MoodyPatera:1993b, DechantTwarockBoehm2011E8A4, Indelicato2011Transitions}.  We therefore briefly outline the basics of such a construction.  

 Let us consider the conformal space of signature $(+,+,+,+,-)$ achieved by adjoining two additional orthogonal unit vectors $e$ and $\bar{e}$ to the algebra of space \cite{HestenesSobczyk1984}. It is therefore spanned by the unit vectors 
\begin{equation}\label{ch0dSbasis}
    e_1, e_2, e_3,  e, \bar{e}, \text{ with } e_i^2=1, e^2=1, \bar{e}^2=-1.
\end{equation}
From these two unit vectors  we can define the  two null vectors
\begin{equation}\label{ch0dSnnbar}
    n \equiv e +\bar{e}, \,\,\,  \bar{n}\equiv e -\bar{e}.
\end{equation}
One can then map a 3D vector $x$ into the space of null vectors in the conformal space by defining 
\begin{equation}\label{ch0nullembedding}
  X\equiv F(x):= x^2 n+2\lambda x-\lambda^2\bar{n}.
\end{equation}
$X$ being null allows for a homogeneous (projective) representation of points, i.e. they are represented by a ray in the conformal space, which  tends to be more numerically robust in applications, as for instance the origin is represented by $\bar{n}$ rather than the number  $0$, which is sensitive to the accumulation of numerical errors. 
Here, $\lambda$ is a fundamental length scale that is needed in order to make this expression dimensionally homogeneous, as we think of the position vector $x$ as a dimensionful quantity \cite{Lasenby2004Acovariantapproach}.
The equivalent notation in terms of the Amsterdam protocol would be $e=e_+$, $\bar{e}=e_-$, $n=n_\infty$ and $\bar{n}=n_0$.
This notation is also consistent with the notion that the above mapping is essentially an embedding into the projective null cone of the embedding space. Originally due to Dirac \cite{Dirac1936}, the idea is that the projective null cone inherits the $SO(4,1)$ invariance of the ambient space in which $SO(4,1)$ acts linearly, thereby endowing the projective null cone with a non-linearly realised conformal structure.

\begin{figure}
\begin{center}
\tikzstyle{background grid}=[draw, black!50,step=.5cm]
    \begin{tabular}{@{}c@{ }c@{ }}
	
				 	\begin{tikzpicture}[scale=1.8,
							    knoten/.style={        circle,      inner sep=.08cm,        draw}  , 
							   dot/.style={        circle,      inner sep=.02cm,        draw} 
							   ]

							  \node at (1, 0)   (hex1) {};
							  \node at (.81, .59) (hex2) {};  
							  \node at (.31, .95) (hex3) {};
							  \node at (-.31, .95) (hex4){};
							  \node at (-.81, .59) (hex5) {};
							  \node at (-1, 0) (hex6)  {};
							\node at (-.81, -.59) (hex7) {};  
							\node at (-.31, -.95) (hex8) {};
							\node at (.31, -.95) (hex9){};
							\node at (.81, -.59) (hex10) {};
							  \node at (0, 0) (Ori)  {};

							  \path [color=gray] (hex2.mid) edge (hex3.mid);
							  \path  [color=gray] (hex1.mid) edge (hex2.mid);
							  \path  [color=gray] (hex3.mid) edge (hex4.mid);
							  \path [color=gray]  (hex4.mid) edge (hex5.mid);
							  \path [color=gray]  (hex5.mid) edge (hex6.mid);
						      \path  [color=gray] (hex1.mid) edge (hex10.mid);
							  \path  [color=gray] (hex6.mid) edge (hex7.mid);
							  \path [color=gray]  (hex7.mid) edge (hex8.mid);
							  \path [color=gray]  (hex8.mid) edge (hex9.mid);
						      \path  [color=gray] (hex9.mid) edge (hex10.mid);

							\path [->] (Ori.mid) edge (hex1.mid);
							\path [->] (Ori.mid) edge (hex2.mid);
							\path [->] (Ori.mid) edge (hex3.mid);
							\path [->] (Ori.mid) edge (hex4.mid);
							\path [->] (Ori.mid) edge (hex5.mid);
							\path [->] (Ori.mid) edge (hex6.mid);
							\path [->] (Ori.mid) edge (hex7.mid);
							\path [->] (Ori.mid) edge (hex8.mid);
							\path [->] (Ori.mid) edge (hex9.mid);
							\path [->] (Ori.mid) edge (hex10.mid);

				\path [->, color=red] (Ori.mid) edge (hex1.mid) ;
				\path [->, color=red] (Ori.mid) edge (hex5.mid) ;


				  \node at (1.2*1, 0)   (nhex1) {{\textcolor{red}{$\alpha_1$}}};
				  \node at (1.4*.81, 1.5*.59) (nhex2) {\scriptsize{$\tau\alpha_1+\alpha_2$}};  
				  \node at (1.2*.31, 1.3*.95) (nhex3) {\scriptsize{$\tau(\alpha_1+\alpha_2)$}};
				  \node at (-1.2*.31, 1.3*.95) (nhex4){\scriptsize{$\alpha_1+\tau\alpha_2$}};
				  \node at (-1.2*.81, 1.3*.59) (nhex5) {{\textcolor{red}{$\alpha_2$}}};
							\end{tikzpicture} &\hspace{0.8cm}

					\begin{tikzpicture}[scale=1.3,
					    knoten/.style={        circle,      inner sep=.08cm,        draw}  , 
					   dot/.style={        circle,      inner sep=.02cm,        draw}  , 
					my node/.style={trapezium, fill=#1!20, draw=#1!75, text=black} 
					   ]

					  \node at (1, 0)   (hex1)  [knoten,  ball color=black] {};
					  \node at (.309, -.951) (hex2) [knoten, ball color=black] {};  
					  \node at (.309, .951) (hex3) [knoten, ball color=black] {};
					  \node at (-.809, -.588) (hex4) [knoten,  ball color=black] {};
					  \node at (-.809, .588) (hex5) [knoten, ball color=black] {};

					  \node at (1+0.618*.309, 0-0.618*.951)   (hex1a)  [knoten,  ball color=gray!30!white] {};
					  \node at (.309+0.618*.309, -.951-0.618*.951) (hex2a) [knoten, ball color=gray!30!white] {};  
					  \node at (.309+0.618*.309, .951-0.618*.951) (hex3a) [knoten, ball color=gray!30!white] {};
					  \node at (-.809+0.618*.309, -.588-0.618*.951) (hex4a) [knoten,  ball color=gray!30!white] {};
					  \node at (-.809+0.618*.309, .588-0.618*.951) (hex5a) [knoten, ball color=gray!30!white] {};

					  \node at (1+0.618*1, 0)   (hex1e)  [knoten,  ball color=gray!30!white] {};
					  \node at (.309+0.618*1, -.951) (hex2e) [knoten, ball color=gray!30!white] {};  
					  \node at (.309+0.618*1, .951) (hex3e) [knoten, ball color=gray!30!white] {};
					  \node at (-.809+0.618*1, -.588) (hex4e) [knoten,  ball color=gray!30!white] {};
					  \node at (-.809+0.618*1, .588) (hex5e) [knoten, ball color=gray!30!white] {};

					  \node at (1+0.618*.309, 0+0.618*.951)   (hex1b)  [knoten,  ball color=gray!30!white] {};
					  \node at (.309+0.618*.309, -.951+0.618*.951) (hex2b) [knoten, ball color=gray!30!white] {};  
					  \node at (.309+0.618*.309, .951+0.618*.951) (hex3b) [knoten, ball color=gray!30!white] {};
					  \node at (-.809+0.618*.309, -.588+0.618*.951) (hex4b) [knoten,  ball color=gray!30!white] {};
					  \node at (-.809+0.618*.309, .588+0.618*.951) (hex5b) [knoten, ball color=gray!30!white] {};

					  \node at (1-0.618*.809, 0-0.618*.588)   (hex1c)  [knoten,  ball color=gray!30!white] {};
					  \node at (.309-0.618*.809, -.951-0.618*.588) (hex2c) [knoten, ball color=gray!30!white] {};  
					  \node at (.309-0.618*.809, .951-0.618*.588) (hex3c) [knoten, ball color=gray!30!white] {};
					  \node at (-.809-0.618*.809, -.588-0.618*.588) (hex4c) [knoten,  ball color=gray!30!white] {};
					  \node at (-.809-0.618*.809, .588-0.618*.588) (hex5c) [knoten, ball color=gray!30!white] {};

					  \node at (1-0.618*.809, 0+0.618*.588)   (hex1d)  [knoten,  ball color=gray!30!white] {};
					  \node at (.309-0.618*.809, -.951+0.618*.588) (hex2d) [knoten, ball color=gray!30!white] {};  
					  \node at (.309-0.618*.809, .951+0.618*.588) (hex3d) [knoten, ball color=gray!30!white] {};
					  \node at (-.809-0.618*.809, -.588+0.618*.588) (hex4d) [knoten,  ball color=gray!30!white] {};
					  \node at (-.809-0.618*.809, .588+0.618*.588) (hex5d) [knoten, ball color=gray!30!white] {};

					\end{tikzpicture}
					
		\\

\end{tabular}	

\end{center}

\caption[dummy1]{In the conformal setup, reflections generated by the simple roots (here e.g. $\alpha_1$ and $\alpha_2$ for a simple two-dimensional example, $H_2$) according to Eq. (\ref{ch0Ereforigin}) again generate, for instance, the $H_2$ non-crystallographic root system, the decagon (left). 
CGA rotor translations via Eq. (\ref{ch0ETransl}) act multiplicatively, but yield quasicrystalline point sets consistent with the 3D approach; for instance, on the right we show the effect of a translation with length the inverse of the golden ratio acting on a pentagon, in analogy to Figs \ref{figpent} and \ref{figpent2}.}
\label{figrootsystem}
\end{figure}
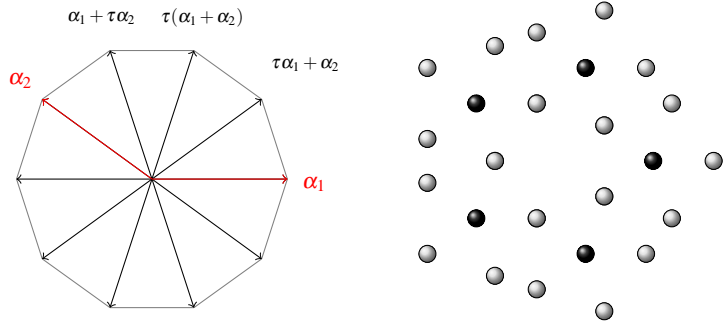

The  vectors $e$ and $\bar{e}$ and therefore also $n$ and $\bar{n}$ are orthogonal to $x$ and hence anticommute with it, i.e. $-x^{-1}nx=n$ and $-x^{-1}\bar{n}x=\bar{n}$. Thus, the CGA implementation of a reflection $y'=-x^{-1}yx$ is given by
\begin{equation}\label{ch0Ereforigin}
  -x^{-1}F(y)x=F(y')=F(-x^{-1}yx).
\end{equation}
Given the simple roots, one can again generate the whole root system via successive reflections as shown in Fig. \ref{figrootsystem} (left). 
We firstly notice that the conformal representation of a root vector $F(\alpha)$ is now different from the implementation of the reflection encoded by it via the versor  $\alpha$. These two roles were treated on an equal footing in 3D, as there $\alpha$ represents both the root vector and the versor encoding a reflection in the hyperplane perpendicular to the root, and it is debatable whether the conceptual advantages of CGA outweigh this disadvantage. 

Secondly, it is often argued that the implementation of rotations in CGA is given by $F(x')=RF(x)\tilde{R}$, since $R$ only contains even blades and thus commutes with the vectors $n$ and $\bar{n}$ such that $Rn\tilde{R}=n$ and $R\bar{n}\tilde{R}=\bar{n}$. However, via the Cartan-Dieudonn\'e theorem, every rotation is given by an even number of successive reflections. Thus, it can be seen that the rotor transformation law actually follows from the more fundamental reflection law in Eq. (\ref{ch0Ereforigin}). From the previous sections, we know that the spinors generated by the root vectors are important for the construction of the binary polyhedral groups and 4D polytopes. However, the 3D geometric product does not straightforwardly extend to CGA, such that the spinors and other multivectors are not treated in the same way as vectors. The operators encoding the conformal rotations, however, are still given by the 3D rotors, so that little seems to be gained by going to the conformal setup from the spinorial point of view.

A very salient feature of Conformal Geometric Algebra is that a translation $x\rightarrow x+a$ by a vector $a$ is given by a rotor
\begin{equation}\label{ch0ETransl}
  T_a= \exp\left(\frac{na}{2\lambda}\right)=1+\frac{na}{2\lambda}.
\end{equation}
It is easily checked that this has  the desired effect of $T_a F(x) \tilde{T}_a=F(x+a)$, and therefore
does indeed represent a 3D translation as a rotor in Conformal Geometric Algebra. One can thus treat reflections, rotations and translations multiplicatively in a unified framework. This allows for a unified construction of the type of point arrays considered earlier, and indeed the construction is entirely equivalent to the lower-dimensional construction (as it must), and can be straightforwardly verified, for instance, for the non-crystallographic groups $I_2(n)$, $H_3$ and $H_4$. In Fig. \ref{figrootsystem}, we show an example consisting of both one such root system and one quasicrystal-like point array derived  entirely in the conformal setup, as a proof of principle. The root system shown is that of $H_2$, and the point array is obtained via the action of a translation of length the inverse of the golden ratio on a pentagon. 

The CGA approach is naturally more computationally intensive than the 3D approach; however, this could be compensated for by increased numerical stability, as the origin is simply represented by scalar multiples of $\bar{n}$, as opposed to the number $0$, where numerical errors can create artefacts near the origin. Treating both rotations and translations on the same footing as multiplicative rotors is also a nice conceptual shift. However, there are also drawbacks  to the conformal approach. Firstly, the conformal representation of the root vectors $F(\alpha)$ is different from their  action as generators of reflections $\alpha$. The relationship between these two functions was more transparent in the conventional approach in 3D, where $\alpha$ represented both. Secondly, the rotors encoding rotations are also the 3D spinors, rather than a conformal representation of those. Thus, CGA affords a nice representation of GA vectors, but not necessarily of the whole GA multivector structure. 

Following \cite{Lasenby2005RecentApplications}, an interesting approach might be to work in a curved space, for which only one extra dimension is necessary ($e$ or $\bar{e}$), which should simplify the computations somewhat. One may then finally take the zero curvature limit in order to recover the Euclidean space results. 
For instance, for Minkowski spacetime, the conformal group $C(1,3)$ is 15-dimensional. It has certain well-known ten-dimensional groups as stabiliser subgroups, i.e. groups of transformations that leave a given point (ray) $y$ invariant. If $y$ is spacelike, one gets an $SO(2,3)$ stabiliser subgroup, i.e.  the Anti de Sitter group, corresponding to the homogeneous spacetime that is the solution of Einstein's field equations with a negative cosmological constant, $\Lambda <0$. Likewise, for timelike $y$ one obtains the de Sitter ($\Lambda>0$) group $SO(1,4)$ as the stabiliser (here in the CGA setup, $e$ and $\bar{e}$ are distinguished choices for such spacelike and timelike $y$). Lastly, when one chooses a null $y$ (e.g. $n$), one gets an $ISO(1,3)$ subgroup, which is just the Poincar\'e group \cite{Porteous1995Clifford, Baugh2004Thesis}. Thus, taking the zero curvature limit essentially corresponds at the group level to the Wigner-In\"on\"u contraction that yields the Poincar\'e group from the de Sitter and Anti de Sitter groups (see e.g. \cite{Freund1988SUSY}) and  a flat space (which needs two vectors $e$ and $\bar{e}$) limit from a curved space (for which only one of $e$ or $\bar{e}$ is necessary).

\section{$I_2(n)$ and $H_3$ -- the Coxeter element and spinors}\label{sec_2D}

In this section, we further analyse the two-dimensional family  of non-crystallographic Coxeter groups $I_2(n)$ (the symmetry groups of the regular polygons), as well as the three-dimensional groups $A_3$, $B_3$ and the icosahedral group $H_3$, describing the symmetries of the Platonic solids. 
A versor framework (not necessarily conformal) allows a deeper understanding of the geometry, relating spinors to the Coxeter element and the Coxeter plane in a novel way, in particular highlighting what the complex structure involved is.

A Coxeter element $w=s_1 \dots s_n$ is the product of the reflections encoded by all the simple roots $\alpha_i$ of a finite Coxeter group $W$. The Coxeter number $h$ is the order (i.e. $w^h=1$) of such a Coxeter element. The sequence in which the simple reflections are performed does matter, but all such elements are conjugate, and thus the Coxeter number $h$ is the same (for instance for $I_2(n)$ one has $h=n$). 
For a given Coxeter element $w$, there is a unique  plane called the Coxeter plane on which  $w$ acts 
as a rotation by $2\pi/h$.
At this point in the standard theory, there is a convoluted argument about the need to complexify the situation and taking real sections of the complexification in order to find the complex eigenvalues $\exp(2\pi i/h)$ and $\exp(2\pi i(h-1)/h)$  \cite{Humphreys1990Coxeter}. 
It will come as no surprise that in Geometric Algebra the complex structure arises naturally, giving a geometric interpretation for the `complex eigenvalues'. 

\begin{figure}
	\begin{center}
	      \begin{tabular}{@{}c@{ }c@{ }c@{ }}
				\begin{tikzpicture}
				\node (img) [inner sep=0pt,above right]
				{\includegraphics[width=3.5cm]{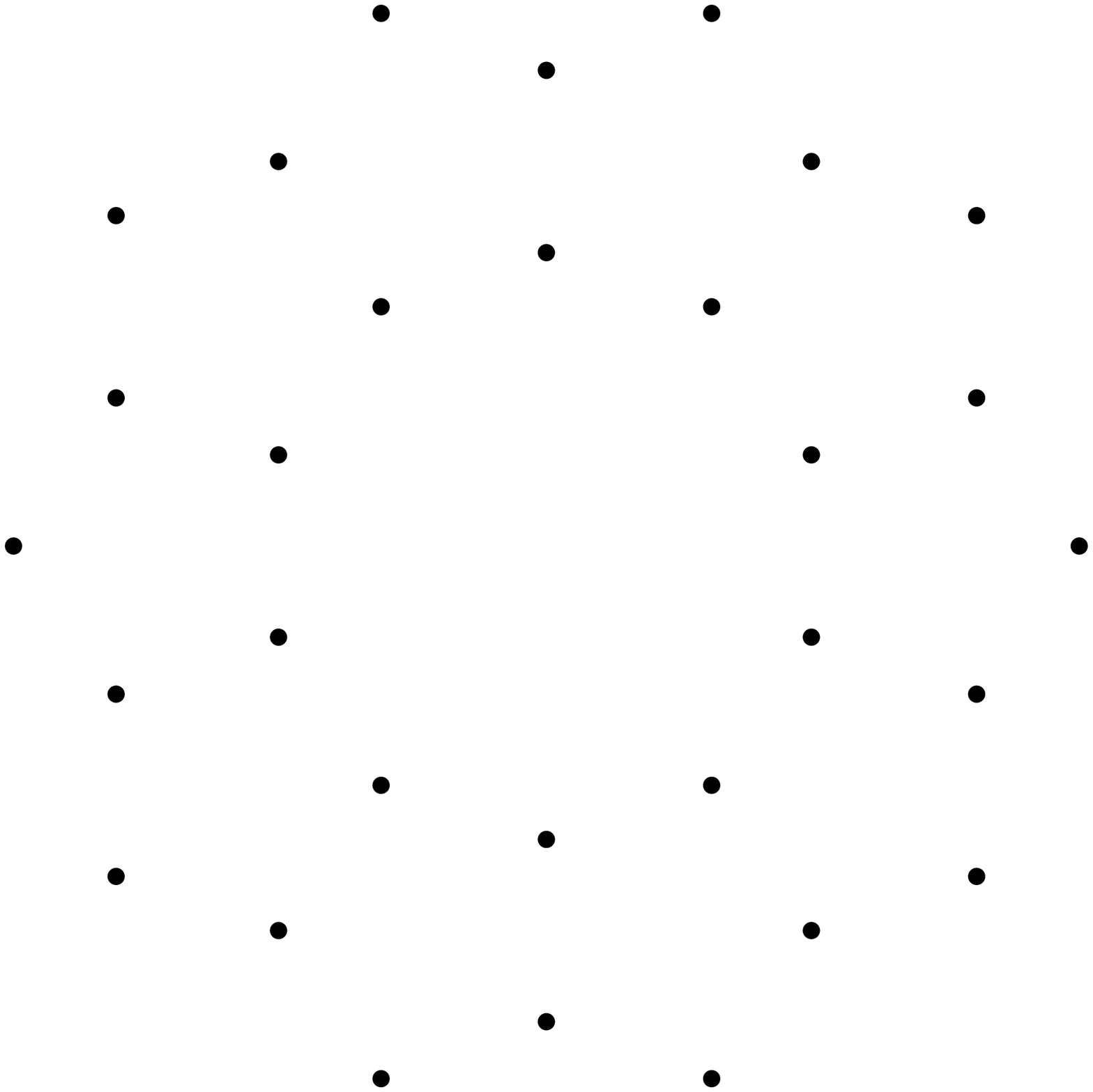}};
				\end{tikzpicture}&\hspace{1.5cm}
				\begin{tikzpicture}
				\node (img) [inner sep=0pt,above right]
				{\includegraphics[width=3.5cm]{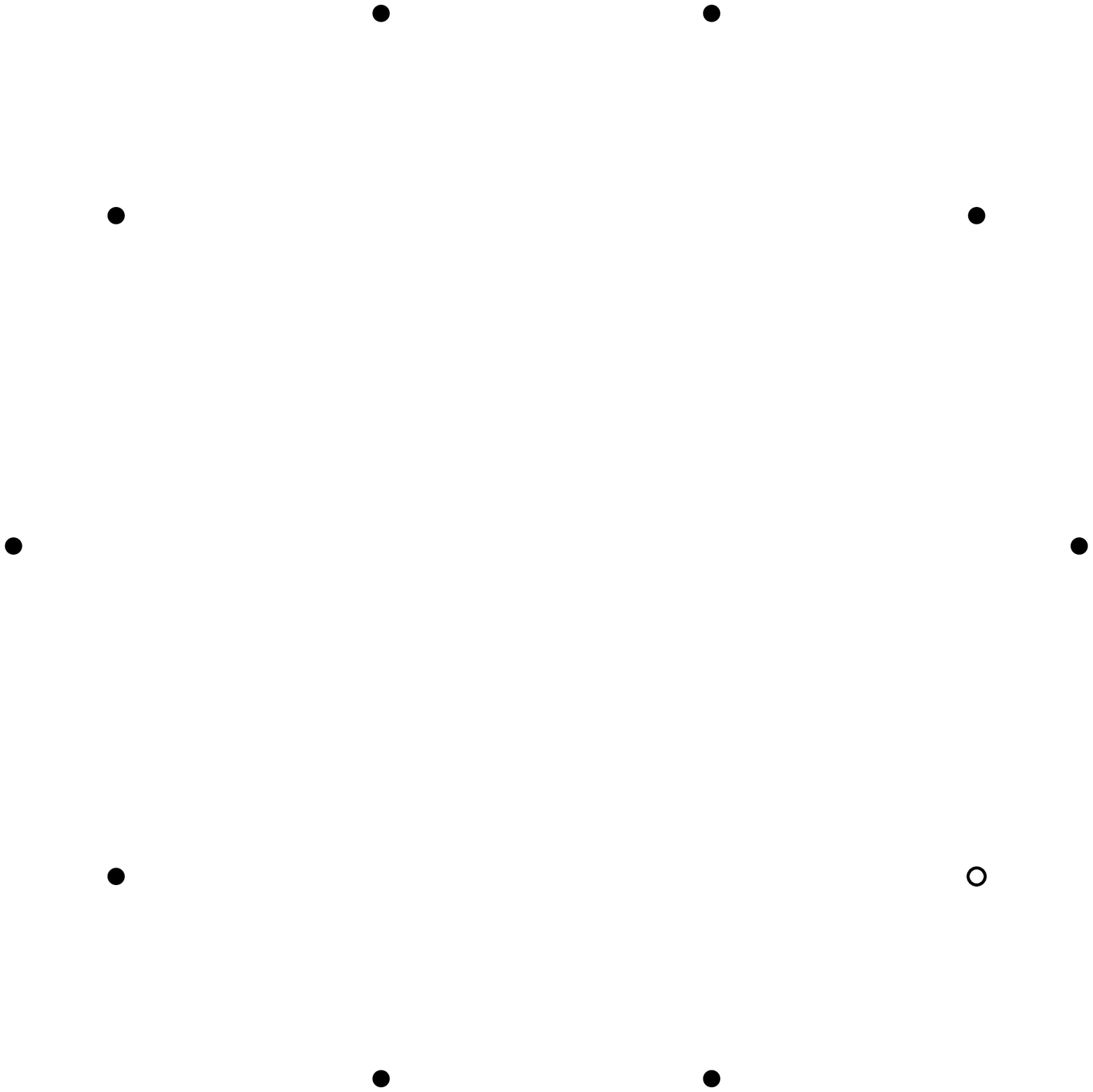}};
			\end{tikzpicture} &\hspace{1.5cm}
				\begin{tikzpicture}
				\node (img) [inner sep=0pt,above right]
				{\includegraphics[width=3.5cm]{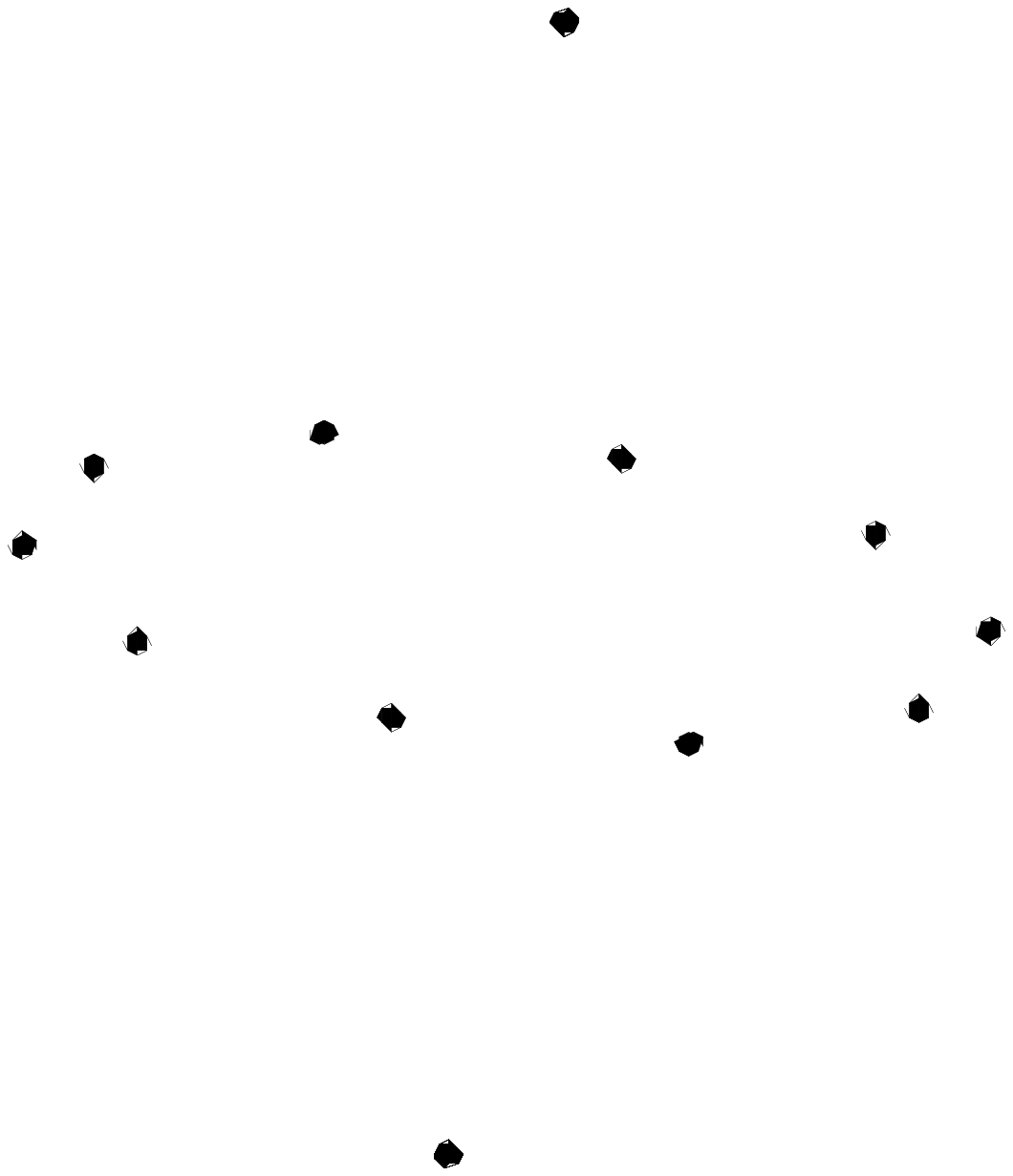}};
				\end{tikzpicture}
		\vspace{0.25cm}
	\\
			(a)&(b) &(c) \\

\end{tabular}	

\end{center}

\caption[dummy1]{Illustration of the geometry of $H_3$. (a) shows the projection of the root polytope (the icosidodecahedron with 30 vertices) onto the Coxeter plane. Panel (b) illustrates the action of the Coxeter element on a vector  $v=e_1$  denoted by the open circle in the Coxeter plane. $w$ acts by 10-fold rotation generating a decagon clockwise, whereas on a  vector $n$ normal to the Coxeter plane it acts by reversal $-n$. Panel (c) displays both sets of vectors, which in turn happen to form the root system of $A_1\times H_2$.  }
\label{figCoxPl}
\end{figure}

 Projection of a root system onto the Coxeter plane is a  way of visualising any finite Coxeter group, for instance the well-known representation of $E_8$ is such a projection of the 240 vertices of the eight-dimensional Gosset root polytope onto the Coxeter plane. Fig. \ref{figCoxPl} (a) shows such a projection of the root polytope of $H_3$ (the icosidodecahedron) onto the Coxeter plane.

Without loss of generality, in Geometric Algebra the simple roots for $I_2(n)$  can  be taken as
 $\alpha_1=e_1$ and $\alpha_2=-\cos{\frac{\pi}{n}}e_1+\sin{\frac{\pi}{n}}e_2$
(see the $H_2$ root system in Fig. \ref{figrootsystem} for $n=5$). 
The Cartan matrix $	A_{ij}=2\alpha_i\cdot \alpha_j/\alpha_i^2$ is then correctly given by  
\begin{equation}
	A(I_2(n)) = \begin{pmatrix} 2&-2\cos{\frac{\pi}{n}} \\ -2\cos{\frac{\pi}{n}}&2 \end{pmatrix}.
\end{equation}

The  Coxeter versor $W$ describing the rotation encoded by the $I_2(n)$ Coxeter element via the typical GA half-angle formula
\begin{equation}\label{halfangle}
	v\rightarrow wv=\tilde{W}vW
\end{equation}
 is therefore 
\begin{equation}
W=\alpha_1\alpha_2=-\cos{\frac{\pi}{n}}+\sin{\frac{\pi}{n}}e_1e_2=-\cos{\frac{\pi}{n}}+\sin{\frac{\pi}{n}}I=-\exp{\left(-\frac{\pi I}{n}\right)}
\end{equation}
 for $I=e_1e_2$. 
In GA it is therefore immediately obvious that the action of the $I_2(n)$ Coxeter element is described by a versor (here a rotor/spinor) that encodes rotations in the $e_1e_2$-Coxeter-plane and  yields $h=n$ since trivially
\begin{equation}
W^n=(-1)^{n+1} \Rightarrow w^n=1.
\end{equation}
More generally, the versors belonging to conjugate Coxeter elements could be $W=\pm\exp{\left(\pm\frac{\pi I}{n}\right)}$ and one immediately finds that $W^n=\pm 1$  as required for $w$ to be of order $h=n$. 

Since $I=e_1e_2$ is the bivector defining the plane of $e_1$ and $e_2$, it anticommutes with both $e_1$ and $e_2$. Thus, in the half-angle formula Eq. (\ref{halfangle}), one can take $W$ through to the left to write the complex eigenvector equation 
\begin{equation}\label{halfangle2}
	v\rightarrow wv=\tilde{W}vW=\tilde{W}^2v=\exp{\left(\pm\frac{2\pi I}{n}\right)}v,
\end{equation}
immediately yielding the standard result for the complex eigenvalues. However, in GA it is now obvious that the complex structure is in fact given by the bivector describing the Coxeter plane (trivial for $I_2(n)$), and that the standard complexification is both unmotivated and unnecessary. The `complex eigenvalues' are simply left and right going spinors in the Coxeter rotation plane.

The Pin group/eigenblade description in GA therefore yields a wealth of novel geometric insight and the general case will be the subject of a future publication. However, for instance for the icosahedral group $H_3$, standard theory has $h=10$ and  complex eigenvalues $\exp(2\pi mi/h)$ with the exponents $m=\lbrace 1, 5, 9\rbrace$. For simple roots $\alpha_1=e_2$, $-2\alpha_2=(\tau-1)e_1+e_2+\tau e_3$ and $\alpha_3=e_3$, one finds the Coxeter plane bivector $B_C=e_1e_2+\tau e_3e_1$.  Under the action of the Coxeter element versor $2W=-\tau e_2-e_3+(\tau-1)I$ (here $I=e_1e_2e_3$) it gets reversed $-\tilde{W}B_CW=-B_C$ as is expected for an invariant plane under an odd operation. For an `eigenvector' in the Coxeter plane, the two-dimensional argument from Eq. (\ref{halfangle2}) applies and one again finds eigenvalues $\exp{\left(\pm\frac{2\pi B_C}{h}\right)}$, which corresponds to $m=1$ and $m=9$. In fact, this holds true for a general Coxeter group: $1$ and $h-1$ are always exponents and in Geometric Algebra they correspond to `eigenvectors' being rotated  in the Coxeter plane via left and right going spinors. However, in Geometric Algebra it is also obvious that in general more complicated geometry is at work, with different complex structures corresponding to different eigenspaces. 
Going back to our $H_3$ example, for the vector $b_C=B_C I=-\tau e_2-e_3$ orthogonal to the Coxeter plane, one has $-\tilde{W}b_CW=-b_C=\exp{\left(\pm\frac{5\cdot 2\pi B_C}{h}\right)}b_C$, as is expected for the normal vector for a plane that gets reversed. Thus, in GA this case straightforwardly corresponds to $m=5$, accounting for the remaining case. 

Fig. \ref{figCoxPl} illustrates this $H_3$ geometry. Panel (a) shows the projection of the root system, the icosidodecahedron, onto the Coxeter plane. The vector $v=e_1$   lies in the Coxeter plane and the Coxeter element $w$ acts on it by 10-fold rotation via the Coxeter versor $W$. This is depicted in Panel (b), where $v$ is denoted by the open circle, and rotation via $W$ occurs in the clockwise direction creating a decagon. The Coxeter versor acts on the vector $b_C$ normal to the Coxeter plane simply by reversal, as discussed above. Both sets of vectors (the decagon and $\pm$ the normal) are depicted in Panel (c). Curiously, these vectors form the root system of $A_1\times H_2=A_1\times I_2(5)$. 

The geometry for $A_3$ and $B_3$ is very similar. They have Coxeter numbers $h=4$ and $h=6$, respectively, and exponents $m=\lbrace 1, 2, 3\rbrace$ and $m=\lbrace 1, 3, 5\rbrace$. The Coxeter versor again inverts the Coxeter bivector, and the exponents $1$ and $h-1$ correspond to left and right going rotations in the Coxeter plane on which the Coxeter element acts by $h$-fold rotation, whilst the normal to the Coxeter plane gets simply inverted as expected, corresponding to the cases $h/2$ ($m=2$ and $m=3$ for $A_3$ and $B_3$, respectively). Again, the combinations of the vectors in the plane and orthogonal to it form the root systems of $A_1\times A_1\times A_1=A_1\times I_2(2)$ and $A_1\times A_2=A_1\times I_2(3)$.


\section{Conclusions}\label{sec_concl}

We have investigated what insight a Geometric Algebra description, which lends itself to applications of reflections, can offer when applied to the Coxeter (reflection) group framework. 
The corresponding computations are conceptually revealing, both for applications to real systems and for purely mathematical considerations. 
The implementation of orthogonal transformations as versors rather than matrices offers some computational and conceptual advantages, in both the conventional and the conformal approaches.
The main benefit in a versor description of the applications, for instance in virology, lies in the simple construction and implementation of the chiral and full polyhedral groups.
The Clifford approach then also yields a simple construction of the binary polyhedral groups, and in fact all three groups can be straightforwardly treated in the same framework.
This seemingly unknown construction of the binary polyhedral groups also sheds light on the fact why they generate Coxeter groups of rank 4.
The natural 4D Euclidean structure of the spinors allows for an alternative interpretation as vectors (in fact, a root system) in a 4D space, which generate Coxeter groups in four dimensions.
Thus, one can construct many four-dimensional (exceptional) Lie and Coxeter groups from three-dimensional considerations alone. 
We have constructed non-crystallographic root systems and groups, as well as quasicrystalline point arrays in the conformal framework. This could be interesting for the latter quasicrystals, as translations (e.g. arising from affine extensions of the Coxeter groups) are treated multiplicatively by versors in the same way as rotations and reflections. 
We have discussed the versor framework for the groups $I_2(n)$, $A_3$, $B_3$ and $H_3$, in particular in relation to the Coxeter element, the Coxeter plane and complex eigenvalues/exponents. The Geometric Algebra approach gives novel geometric insight, as the complex structure is seen to arise from the Coxeter plane bivector, and the Coxeter element acts as a spinor generating rotations in this Coxeter plane. 

We are currently applying the more formal considerations of our recent work to extending the existing paradigm for modeling virus and fullerene structure \cite{DechantTwarockBoehm2011Jess} and to packing problems \cite{KeefDechantTwarock2012Packings}. The chiral and binary polyhedral groups are attractive as discrete symmetry groups for flavour and neutrino model building in particle physics, and we are currently working on an anomaly analysis (breaking of classical symmetries by quantum effects) for these groups \cite{Dechant2012Anomalies}. The two-dimensional groups $I_2(n)$ generate the symmetries of protein oligomers, which we are currently investigating.

\begin{acknowledgement}
I would like to thank my family and friends for their support, my former PhD supervisor Anthony Lasenby for getting me interested in  and  teaching me GA (amongst many other things),  as well as David Hestenes,  Eckhard Hitzer, Joan Lasenby, Mike Hobson, Reidun Twarock,  C\'eline B\oe hm, Christoph Luhn and Silvia Pascoli for helpful discussions.
Many thanks also to the referees, who have led to many valuable improvements in the manuscript.
\end{acknowledgement}
\bibliography{/users/pierre-philippedechant/Dropbox/work_share/virus,/YCSSA/virobib}

\end{document}